\documentclass{article}

\usepackage{amsmath, amssymb}
\usepackage{graphicx}
\begin{document}
\title{Dynamics of Spin-Polarized Electron Liquid and Spin Pendulum}
\author{R.N.Gurzhi, A.N.Kalinenko, A.I.Kopeliovich, P.V.Pyshkin, A.V.Yanovsky}

\begin{abstract}
We investigate the dynamics of spin-nonequilibrium electron
systems for the case when normal electron collisions prevail over
the other scattering processes and the "hydrodynamic flow" regime
is realized. The hydrodynamic equations for the electron liquid
have been obtained and analyzed. We demonstrate that oscillations
of the spin polarization are possible in a conducting ring with
inhomogeneous magnetic properties. These low-decay oscillations
are accompanied by the oscillations of the drift current in the
ring. We demonstrate also that the spin polarization of the
electron density may be revealed via the voltage between the ends
of the open circuit with an inhomogeneous spin polarization. The
effect may be observed both in the hydrodynamic and diffusive
regimes.
\end{abstract}

\maketitle

{\bf{Introduction.}} Last years, generation of nonequilibrium
spin-polarized electron states in non-magnetic conductors and
manipulation with them were in the focus of attention (see, e.g.
\cite{Zutic}). A number of promising methods and schemes was
proposed and discussed; the most of them operate under the
conditions when electron-electron interactions are neglected.
Effects of electron-electron scattering in the spin dynamics were
considered in Refs. 2-4, but it was not for the case of
hydrodynamic electron flow \cite{gur1}, when the electron-electron
scattering dominates and the electron system may be considered as
a liquid with its' inherent effects.

Meanwhile, a hydrodynamic flow regime is quite real in a
low-dimensional electron gas in heterostructures \cite{dav1} (as
well as, in electron systems over the liquid helium surface
\cite{bunt1}). The main condition is the following. The transport
electron mean free path should be determined by the "normal"
collisions that conserve the total momentum of the system of
interacting particles. It may be electron-electron collisions
(when the Umklapp processes are absent due to small sizes of the
Fermi surfaces of low dimensional conductors) or electron-phonon
collisions (when phonons are tightly coupled to the electron
system).  In other words, the condition  $l_N << l_V$ should be
satisfied, where $l_N$ is the mean free path (m.f.p.) for the
normal collisions, $l_V$ is m.f.p. for bulk collisions that do
not conserve the quasi-momentum.

A hydrodynamic electron flow was observed experimentally in a
high-mobility electron gas in (Al,Ga)As heterostructures
\cite{molen1} (see, also Ref .\cite{gur2}) at 1,5 - 20 K.  For
that range of the electron temperatures the electron-electron
m.f.p., $l_{ee}$, is much less than the electron m.f.p. determined
by the collisions with imperfections of the heterostructure,
$l_i$. Beside that, in the experiment \cite{molen1} the phonon
temperature was much lower than the electron temperature and
collisions with phonons were unessential. The hydrodynamic regime
breaks under increase of the temperature of the sample, because
the electron-phonon m.f.p., $l_{ep}$, becomes less than $l_{ee}$,
and phonons can remove effectively a momentum from the
low-dimensional electron system being in a good acoustic contact
with the surrounding media. If, in the experiment, it is possible
to provide a weak contact with the surrounding media (that results
in the conservation of the total momentum of the electron-phonon
system), then the increasing of the temperature favors the
hydrodynamic regime due to the decreasing of the normal m.f.p.,
$l_N \approx (l_{ee}^{-1} + l_{ep}^{-1})^{-1}$.

In the hydrodynamic flow regime, the state of the electron liquid
is characterized by the velocity $\mathbf u(\mathbf r)$ and its
density $\rho(\mathbf{r})$. Naturally, when our electron system is
spin-polarized, the densities of the spin components differ from
each other. In this case the electron liquid is a two-component
mixture and the density variables have the spin indexes, i.e.
$\rho_\sigma(\mathbf r)$.  Meanwhile, in the leading
approximation, the velocity $\mathbf u(\mathbf r)$ is the same for
the all spin components. The reason is that frequent collisions
between electrons with different spins form a common drift of the
electron system. Moreover, we have to regard the electron liquid
as incompressible when the geometric size of a conductor is
larger than the electron screening radius (which is comparatively
small in metals and heterostructures) and the characteristic
frequencies of the considered processes are less than the plasma
frequency. We would like to emphasize that the incompressibility
means here that the nonequilibrium addition to the electron
density is vanishing (nevertheless, the equilibrium density may
be inhomogeneous due to the reasons that are discussed below).
Obviously, the incompressibility and the equal velocities of the
spin components lead to the following simple fact. The total
current through the channel cross-section, $I$, does not depend on
the coordinate along the channel and it is distributed between
the spin components in proportion to their densities. Thus, the
current characteristics of the system are determined by the value
of $I$.

As we demonstrate below, in the conducting ring with
inhomogeneous magnetic properties the value of the total current
undergoes low-decay harmonic oscillations which frequency depends
on the characteristics of the inhomogeneity. The nature of these
oscillations is the following. The drift in the
spin-inhomogeneous ring causes appearing of a non-equilibrium
spin polarization, i.e. an accumulation of the non-equilibrium
densities of the spin-up and spin-down components occurs (while
the total density is conserved). The accumulation exists until
the moment when the drift is stopped due to the interaction of
non-equilibrium electrons with a field that induces the
inhomogeneity of the electron spectrum. But, electrons have the
inertial masses and the process will evolve back. We named this
oscillation process the "spin pendulum".

Note, the existence of the well-known hydrodynamic waves, which
frequencies are less than the plasma frequency, is impossible here
due to the Coulomb interaction. Consequently,  "spin
pendulum-like" oscillations are the only possible oscillations of
the system in this case.

The interesting spin-electrical effect related to the
non-equilibrium spin polarization may be observed not only in the
ring but in the open circuit as well. The skewed spin
polarization causes a voltage between the ends of the open
circuit and it allows easily registering of the spin
polarization. Note, the spin-electrical effect in magnetoelectric
materials in the equilibrium inhomogeneous state was discussed
early in \cite{smol1}.

Note that the all aforementioned effects are possible both in the
magnetic conductors, where densities of the spin components
differs from each other initially (e.g., due to the exchange
interaction), and in non-magnetic materials, where either spin
separation is due to the Rashba effect \cite{rash1}, or it is
caused by an external magnetic field. It is important that
non-magnetic materials get magnetic properties due to the
appearing of the induced non-equilibrium spin distribution.
Consequently, the aforementioned steady-state and dynamic effects
exist in them but it is the second-order effects in the spin
non-equilibrium.  In the case of a two-dimensional electron gas
in heterostructures, the variation of magnetic properties can be
induced by a non-uniform gate voltage applied, which affects the
Rashba effect \cite{nitta}; by a variation of the external
magnetic field or by a space-dependent spin injection.

{\bf{Spin hydrodynamics}} A time-dependent distribution function
$f(\mathbf {r,p},t)$ for electrons at the position $\mathbf r$ and
with the momentum $\mathbf p$ obeys the Boltzmann transport
equation
\begin{equation}\label{eq1}
    \frac{\displaystyle\partial f}{\displaystyle\partial t} +
    \mathbf v \frac{\displaystyle\partial f}{\displaystyle\partial\mathbf r} =
    \frac{\displaystyle\partial(\varepsilon + e\varphi)}{\displaystyle\partial\mathbf r}
    \frac{\displaystyle\partial f}{\displaystyle\partial\mathbf p} +
    J(f).
\end{equation}
The hydrodynamic equations can be derived from the Boltzmann
transport equation by the method used in Refs. \cite{gur3, gur4},
taking into account that the function $f(\mathbf {r,p},t)$ depends
also on the spin index $\sigma$, which corresponds to the
different spin components. It is assumed in Eq.(\ref{eq1}) that
the energy spectrum of the current carriers depends on the
coordinates, momentum and spin index:
$\varepsilon=\varepsilon(\mathbf r, \mathbf p, \sigma)$. The
electrical potential $\varphi$ appears due to the space-dependent
non-equilibrium spin density. $J(f)$ is the scattering term and
$\mathbf v =\partial\varepsilon/\partial\mathbf p$ is the electron
velocity.

Let us assume that momentum dissipation is vanishing and the
normal scattering processes dominate, and the following
conditions are fulfilled: $l_N/L<<1$ and $\omega\tau_N << 1$.
Here $L$ is the characteristic geometric size of the system,
$\omega$ is the characteristic frequency of the oscillation
processes in the system, $\tau_N=l_N/v_F$ is the relaxation time
corresponding to the normal collisions, and $v_F$ is the Fermi
velocity. In this case we may expand Eq.(1) in series in these
small parameters. In the leading approximation we obtain the
following equation
\begin{equation}\label{eq2}
J_N(f^{(0)})=0.
\end{equation}
Here $J_N$ is the part of the collision term which corresponds to
the normal collisions. The solution of Eq.(\ref{eq2}) is the
quasi-equilibrium drift distribution
\begin{equation}\label{eq3}
f^{(0)} = n(\varepsilon_\sigma - \delta\mu_\sigma -
\textbf{up}),\; n(z) = (e^{(z - \mu)/T} + 1)^{-1},
\end{equation}
where the drift velocity $\mathbf u$  and nonequilibrium additions
to the chemical potential $\delta\mu_\sigma$ depend on the
coordinates $\mathbf r$ and time $t$ (we neglect here the
temperature variations). Following the method used in Refs.
\cite{gur3, gur4} we analyzed the conditions of solvability up to
the next two orders of series expansions and have obtained the
following equations for $\delta\mu_\sigma, \mathbf u$, and
$\varphi$
$$
\left \{
\begin{aligned}
\frac{\partial\delta\rho_\sigma}{\partial t} + \mathrm{div}
{\rho_0}_\sigma\mathbf{u} = (\hat{D} + \hat{F})\delta\rho_\sigma,\label{eq4}\\
m\rho\frac{\partial\mathbf{u}}{\partial t} +
\sum\limits_\sigma\rho_\sigma \nabla(\delta\mu_\sigma + e\varphi)
= (\hat{V} + \hat{U})\mathbf{u},\label{eq5}\\
\sum\limits_\sigma\delta\rho_\sigma = 0, \label{eq6}\\
\rho_\sigma = \rho_\sigma(\mu_\sigma).
\end{aligned}
\right.
$$
Here $\rho_\sigma = {\rho_0}_\sigma + \delta\rho_\sigma$ is the
spin-dependent current density, and ${r_0}_\sigma$ is the
equilibrium spin density. (Note that $\delta\rho_\sigma$ is the
function on $\delta\mu_\sigma$, and, in the linear approximation,
$\delta\rho_\sigma = \Pi_\sigma \delta\mu_\sigma$, where
$\Pi_\sigma$ is the spin-dependent density of states on the Fermi
surface.) $m\rho =
\left(\frac1h\right)^rr^{-1}\sum\limits_\sigma\int
p^2\left(-\frac{\partial n}{\partial\varepsilon}\right)d^rp$,
$\rho = \sum\limits_\sigma\rho_\sigma$, where r is the
dimensionality of the electron system. $\hat D$ is the diffusion
operator; $\hat F$ is the spin-flip operator; the operator $\hat
V$ relates to the viscosity and the operator $\hat U$ corresponds
to the scattering with momentum dissipation (see, e.g. \cite{gur3,
gur4}). Note that the specific forms of these operators are not
important for us: the corresponding terms in
Eqs.(\ref{eq4}-\ref{eq5}) are small enough and we use these terms
only for the estimates.

Meanwhile, Eqs.(\ref{eq4}- \ref{eq6})  have a clear physical
meaning. Equation (\ref{eq4}) describes the law of conservation of
the number of colliding particles, while equation (\ref{eq5})
corresponds to the conservation of the total momentum. Equation
(\ref{eq6}) is the result of incompressibility of the electron
liquid with the Coulomb interaction. (To derive equations (4-5) in
the form presented, we assume that electrons have a spherical
Fermi surface.) Eqs. (\ref{eq4}-\ref{eq5}) are written in the
linear approximation in the drift velocity $\mathbf u$, but the
terms that contain $\delta\mu_\sigma$, $\delta\rho_\sigma$ are
given in the explicit form under the assumption
$\mathbf{u}\rightarrow 0$. This approximation is sufficient here
since the non-linear effects will be considered for the case when
the common drift is absent.

Neglecting both the viscosity of the electron liquid and momentum
dissipation we rewrite Eq.(\ref{eq4}) in the following form
($\delta n = n(\varepsilon-\delta\mu_\sigma) - n(\varepsilon)$)
\begin{align}
m\rho\frac{\partial\mathbf{u}}{\partial t} + \nabla P +
e\rho\nabla\varphi+\sum\limits_\sigma\int\frac{\nabla\varepsilon_\sigma\delta
n}
{h^r} d^rp = 0, \label{eq7}\\
P = \sum\limits_\sigma\int \frac{\mathbf{p}\,\mathbf{v}\,\delta
n}{rh^r} \,d^rp , \quad \frac{\displaystyle\partial
P}{\displaystyle\partial\delta\mu_\sigma} = \rho_\sigma.
\label{eq8}
\end{align}
Note that, integrating Eq.(\ref{eq7}) by parts in the momentum
space, we obtain Eq.(\ref{eq5}). Equation (\ref{eq7}) is a natural
generalization of the Euler equation \cite{landau} for the mixture
of liquids interacting with an external electrical field and with
the field that induces the inhomogeneity of the electron spectrum.
$P(\delta\mu_\sigma)$ is a non-equilibrium addition to the
electron pressure.

In the case of a conducting ring, the electron pressure does not
cause the net effect on the electrons, this is because
$\oint\nabla P dx = 0 $(if our conductor is homogeneous, pressure
can be compensated by an electrical field, $\mathbf{\nabla}
\varphi$). Meanwhile, the force $\mathbf{\nabla}
\varepsilon_\sigma$ may cause a common drift in the ring as it
acts on the electrons which are non-equilibrium on the spin.
Figure 1 illustrates the physical picture arising in the
conducting ring with the inhomogeneity of the equilibrium spin
density and explains the nature of the total force añting on the
spin-nonequilibrium distribution caused by the electron drift
along the inhomogeneous ring.

\begin{figure}
\includegraphics{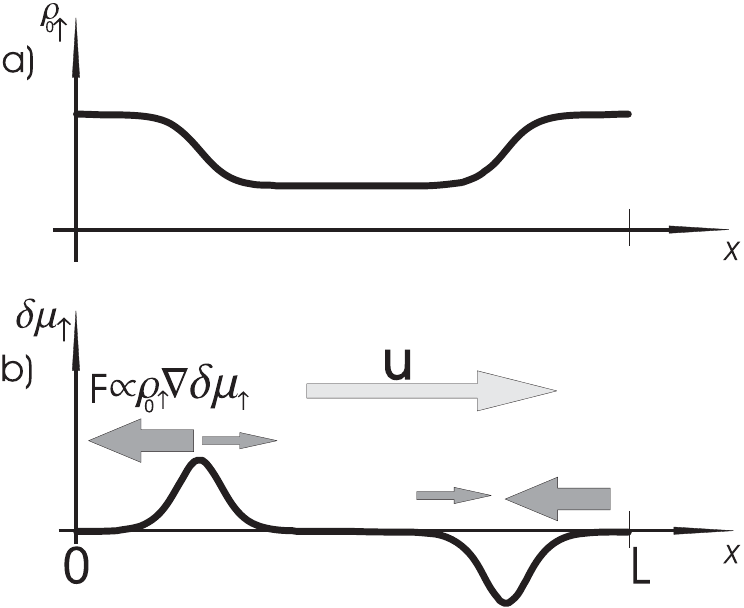}
\caption{Equilibrium distribution of the spin density
${\rho_0}_\uparrow$ along the ring of length  $L$ (Fig.1a). When
electrons drift into the right (see Fig. 1b) the non-equilibrium
spin-up potential $\delta\mu_\uparrow$ appears in the regions
where the spin-up density ${\rho_0}_\uparrow$ varies. The total
force is a sum of all forces acting on electrons and it is
directed to the left. Consequently, it leads to the change of the
drift direction.} \label{f1}
\end{figure}

{\bf{Spin pendulum.}} Let us discuss the oscillation process
described by Eqs. (\ref{eq4}-\ref{eq6}) which is related to the
current $I = I_0e^{i\omega t}$, i.e. a spin pendulum.  In the
linear approximation we may solve the problem completely and find
the frequency of the oscillations
\begin{equation}\label{eq9}
\omega^2 =
{\displaystyle\oint{\left(\frac{\rho_\uparrow}{\rho}\right)'}^2
\frac{1}{\Pi^*}dx}\left(\displaystyle\oint\frac{m}{\rho}dx
\right)^{-1}, \; \frac{\displaystyle 1}{\displaystyle \Pi^*}
=\sum\limits_\sigma\frac{\displaystyle 1}{\displaystyle
\Pi_\sigma}.
\end{equation}
\begin{equation}\label{eq10}
\begin{aligned}
 u=\frac{\displaystyle I}{\displaystyle es\rho}, \quad
\delta\mu_\sigma = i\frac{\displaystyle I}{\displaystyle
es\omega\Pi_\sigma}\left(\frac{\displaystyle\rho_\sigma}{\displaystyle\rho}\right)', \hphantom{owfjeoiej}\\
\quad e\varphi' = -\frac{i}{\omega es}\left[\frac{\displaystyle
m\omega^2}{\displaystyle\rho} +
\sum\limits_\sigma\frac{\displaystyle\rho_\sigma}{\displaystyle\rho}\left(\frac{\displaystyle
1}{\displaystyle\Pi_\sigma}\left(\frac{\displaystyle\rho_\sigma}{\displaystyle\rho}\right)'\right)'\right]I
\end{aligned}
\end{equation}
Here $x$ is the coordinate along the conducting channel, "up
arrow" corresponds to the one of spin components; $s$ is the
cross-section of the channel which is assumed $x$-independent;
strike denotes differentiation with respect to $x$, $L$ is the
length of the conducting channel (i.e. the ring length, in our
case).

As is evident from Eqs.(\ref{eq9}-\ref{eq10}), the oscillations
exist only when the relative concentration of the spin components
varies along the channel. The frequency of oscillations is
proportional to the level of the magnetic inhomogeneity, $\alpha =
({\rho_\sigma}_{max} - {\rho_\sigma}_{min})/ \rho$, viz.
\begin{equation}\label{eq11}
\omega\approx\alpha{\displaystyle v_F}{\displaystyle(LL_1)}^{-1},
\end{equation}
where $L_1$ is the characteristic scale of the spectrum
inhomogeneity.

Initially, the oscillations of the spin-pendulum could be exited
by the "magnetic push" when an external magnetic field is
switched on (much faster than $\omega^{-1}$) and an initial
current $I = es\Phi/cm$ is induced (where $\Phi$ is the magnetic
flux through the ring). Naturally, there are dissipation
processes in the hydrodynamic flow, which we did not take into
account. They lead to the damping of the spin-pendulum
oscillations. It is easy to estimate the decay time
\begin{equation}\label{eq12}
\tau_d\approx\left(\frac{e^2\rho sR}{mL} + \frac{D}{L_1^2} +
\frac{1}{\tau_{sf}}\right)^{-1},
\end{equation}
where $R$ is the electrical resistance of the homogeneous ring.
The second term in the r.h.s. of Eq.(\ref{eq12}) is related to the
diffusion of the nonequilibrium spins through the inhomogeneous
region. In the hydrodynamic regime the diffusion coefficient is
determined by the normal collisions, i.e. $D\approx l_{N}v_F$. The
third term in the r.h.s. of Eq.(\ref{eq12}) corresponds the spin
relaxation which is due to the spin-flip with the characteristic
time  $\tau_{sf}$. Note, as it follows from Eq.(\ref{eq12}), the
"space-sharp" inhomogeneity (e.g., a sharp interface boundary
between different magnetic materials) causes a very fast decay of
the oscillations.

It seems interesting that there exists a possibility to keep the
amplitude of the oscillations constant (like in an ordinary
pendulum clock) owing to the magnetic connection of the ring with
an energy-pumping cell.

{\bf{Spin-electrical effect.}} Let us discuss now a
spin-electrical effect in an open-ended homogeneous conductor. At
${\mathbf u} = 0$ we obtain from Eqs. (\ref{eq7}) - (\ref{eq10})
the following potential difference between the open ends
\begin{flalign}
\varphi(L) &- \varphi(0) =  \frac{e}{\rho}[P(0) - P(L)],&&
\label{eq16}\\
P(x) &= \left(\rho_\uparrow -
\frac{\rho_\downarrow\Pi_\uparrow}{\Pi_\downarrow}\right)\delta\mu_\uparrow
+  \left(\Pi_\uparrow - \frac{\rho_\uparrow}{\Pi_\uparrow}
\frac{\partial\Pi_\uparrow}{\partial\mu_\uparrow}\right)
\delta\mu_\uparrow^2+...&& \label{eq17}
\end{flalign}
 Eq. (\ref{eq17}) yields the electron pressure as the expansion in
series on the $\delta\mu$. The second order term is written in the
form for the case of non-magnetic materials. All the quantities in
the coefficients of $\delta\mu_\uparrow$ and
$(\delta\mu_\uparrow)^2$ are understood as the equilibrium state
quantities. It is obvious from Eq.(\ref{eq16}) that the difference
between the non-equilibrium spin pressures at the ends of the
circuit induces an electrical voltage that can be measured
experimentally. The voltage exists during the lifetime
$\tau_\varphi$ which is determined either by the spin relaxation
due to the spin-flip processes or by the diffusion equalization of
the spin concentration along the circuit, $\tau_\varphi\approx
(\tau_{sf}^{-1} + D/L^2)^{-1}$. The electrical charge $q\approx
\Delta\varphi\tau_\varphi$ flows through the meter during this
time. Let us discuss the case when the spin polarization is
induced in a non-magnetic conductor due to the current spin
injection from the magnetic material. Then, a voltmeter connected
to the ends of the non-magnetic conductor will fix the electrical
charge $q$ and indicate the non-equilibrium spin density be
existed in the non-magnetic conductor.

We should note here that this "spin-electrical" effect is not a
specifically hydrodynamic effect. In the diffusion regime, when
collisions that do not conserve momenta are dominated over normal
collisions, we may write the continuity equations in the
following form
\begin{equation}\label{eq18}
\frac{\displaystyle\partial\delta\rho_\sigma}{\displaystyle\partial
t } +
\mathrm{div}\sigma_\sigma(\nabla(\delta\mu_\sigma+e\varphi))=0,
\end{equation}
where $\sigma_\sigma$ is the contribution  to the conductivity of
the corresponding spin component, $\sigma =
\sum\limits_\sigma\sigma_\sigma$. Consequently, taking into
account Eq.(\ref{eq6}) we obtain the following equation for the
case of homogeneous conductor
\begin{multline}
e\varphi(x)=\left(\frac{\sigma_\downarrow\Pi_\uparrow}{\Pi_\downarrow}
- \sigma_\uparrow\right)\frac{\delta\mu_\uparrow}{\sigma} +\\
\left(\frac{\sigma_\uparrow}{\Pi_\uparrow}\frac{\partial\Pi_\uparrow}{\partial\mu_\uparrow}
-\frac{\partial\sigma_\uparrow}{\partial\mu_\uparrow}\right)\frac{\delta\mu_\uparrow^2}{\sigma}+\ldots
\end{multline}
Here, the second term in the r.h.s. of Eq.(19) is written for the
case of non-magnetic conductors only.

{\bf{Conclusion}}. In summary, the low-decay oscillations of the
spin polarization with the frequency $\omega$ accompanied by the
oscillations of the drift current may be induced in the conducting
ring with inhomogeneous magnetic properties under the condition
$\tau_N << \omega^{-1} << \tau_d$ (see Eq.(\ref{eq9}, \ref{eq12}).
In the case of the heterostructures, the inhomogeneity may be
induced by the inhomogeneous Rashba splitting due to the
space-dependent electrical field. Assuming that the splitting
$\Delta p_F$ is small enough, i.e. $\Delta p_F /p_F << 1$, we
obtain from Eq.(\ref{eq12}) that \mbox{$\omega = v_F
[\frac{2}{L}\oint{(\frac{\Delta p_F}{p_F})'}^2dx]^{1/2}$}. In the
case of a completely nonmagnetic and inhomogeneous ring the
excitation of local spin polarization also induces non-linear
"spin-pendulum-like" oscillations, but the consideration of that
goes beyond the framework of this paper. In the case of the open
circuit, one may detect a non-equilibrium spin-polarization,
measuring the voltage between the open ends of the circuit. It can
be done both in the hydrodynamic and diffusion regimes.

{\bf{Acknowledgements.}} The work was supported in part by
NanoProgram of the NAS of Ukraine (Grant No. 3-026/2004) and by
the joint Ukraine - Byelorussia Grant No. 10.01.006.

\end{document}